\documentclass[letterpaper,12point,epsfig]{article}
\usepackage{epsfig}

\begin{document}
\begin{title}
\title
\begin{center}
\huge{Increasing the Size of a}

\huge{Piece of Popcorn}

\end{center}
\vspace{.2 cm}
\begin{center}
\large{Paul V. Quinn Sr.}

\small{Department of Physical Sciences, Kutztown University, Kutztown, 
Pennsylvania 19530} 

\vspace {.2 cm}

\large{Daniel C. Hong}

\small{Physics, Lewis Laboratory, Lehigh University, Bethlehem, Pennsylvania 18015}

\vspace {.2 cm}

\large{J. A. Both}

\small{Department of Radiation Oncology, Stanford School of Medicine, 
300 Pasteur Drive, Stanford, California 94305}
\end{center}
\date{today}
\end{title}
\begin{abstract}

Popcorn is an extremely popular snack food in the world today. Thermodynamics 
can be used to analyze how popcorn is produced. By treating the popping mechanism 
of the corn as a thermodynamic expansion, a method of increasing the volume or 
size of a kernel of popcorn can be studied. By lowering the pressure surrounding 
the unpopped kernel, one can use a thermodynamic argument to show that the expanded 
volume of the kernel when it pops must increase. In this project, a variety of 
experiments are run to test the validity of this theory. The results show that there 
is a significant increase in the average kernel size when the pressure of the 
surroundings is reduced.

\end{abstract}

\section{Introduction}
Popcorn is an extremely popular snack food, found in many cultures
throughout the world. It was first grown by the American Indians,
before Columbus discovered America. The mass production of popcorn
has become a thriving industry in our society today. Because of the
desire to maximize profits, the food industry may have a vested
interest in trying to increase the size of a piece of popcorn. Larger
popcorn could result in costs savings for mass production as well as 
creating a snack that is more esthetically pleasing to the consumer.
Therefore, larger popcorn would be profitable to the manufacturers as
well as more enjoyable for the consumer. Three major parameters of interest 
to industry are the expansion volume, the flake size, and the percent of
unpopped kernels. We respectively assign the
following variables to these three parameters with their proper
definitions [1,2]:

$$\sigma = {\mathrm {\frac{total\hspace{.2cm} popped\hspace{.2cm}
volume\hspace{.05cm} (cm^3)}{original \hspace{.2cm}sample
\hspace{.2cm}weight \hspace{.05cm}(g)}}},$$
$$\pi = {\mathrm {\frac{total \hspace{.2cm}popped \hspace{.2cm}volume
\hspace{.05cm} (cm^3)}{number \hspace{.2cm}of \hspace{.2cm}popped
\hspace{.2cm}kernels}}},$$
$$\omega = {\mathrm {\frac{number \hspace{.2cm}of
\hspace{.2cm}unpopped \hspace{.2cm} kernels}{original
\hspace{.2cm}number \hspace{.2cm}of \hspace{.2cm}kernels}}}.$$

\noindent In popcorn production, the goal is to make $\sigma$ and $\pi$ as 
large as possible, while trying to minimize $\omega$. To date, the
values $\sigma \approx 45$ cm$^3$/g and $\pi \approx 8$ cm$^3$/kernel,
produced by both an air popper and the microwave, are the largest
expansion volume and flake size achieved by industry. These numbers
are very high and can only be observed under ideal experimental
conditions. The average values achieved by the everyday consumer are
$\sigma \approx 36 - 40$ cm$^3$/g and $\pi \approx 5-7$ cm$^3$/kernel
[1]. Furthermore, the lowest observed percentage of unpopped
kernels is $\omega \approx 6.8 \%$. Once again this is observed under
ideal laboratory conditions, with the consumer finding $\omega \approx
10-12 \%$. [1] Obtaining limits better than those reported by
industry and food scientists by using a more efficient method of popping
popcorn would be of great interest to popcorn producers. There are
many other properties of popcorn that are of interest to industry such
as sphericity, pericarp or shell thickness, internal temperature, and
moisture content. However, an analysis of these characteristics would
be mathematically complex, and not as practical for increasing popcorn
production [3,4]. The purpose of this chapter is to analyze
the physics of popcorn production, using some fairly simple
thermodynamic principles, and to predict a method for controlling or
increasing the size of the popped popcorn. We then compare the results of
the theory with various experiments. Then we discuss our results and their 
application to the popcorn industry.

\section{Thermodynamics of Popping Popcorn}
Before discussing an increase in the size of a popped kernel, we must examine 
the general physics behind the production of popcorn. A popcorn kernel 
consists of starch surrounded by a hard exterior shell called the pericarp. 
Inside every unpopped kernel of popcorn, there is moisture trapped among the 
starch. When moisture is heated, it is prone to expand. As the popcorn 
is heated, the starch inside develops a jelly-like consistency, allowing 
the moisture to uniformly mix with the starch. When the temperature 
of the heated kernel exceeds the boiling point of water, the moisture inside 
changes to a gas, causing a sudden expansion in volume. However, the hard 
shell or pericarp, which coats the outside of the kernel, keeps the gas trapped 
inside the shell. Hence, the internal pressure of the kernel rises as 
the popcorn continues to be heated. When the internal pressure of the kernel 
reaches the yield or breaking point of the shell, the pericarp splits, and the 
water vapor inside rapidly expands, taking the jelly-like starch with it. The 
rapid expansion into the surrounding air cools the starch, forming the fluffy 
substance known as popcorn. This expansion can be modeled as adiabatic 
because the speed of the expansion does not allow the gas to exchange heat and 
reach an equilibrium with the environment. Rather, the expansion stops when 
the pressure of the gas reaches the pressure of the air surrounding the 
kernel. This is why freshly popped popcorn is still hot. For more details on 
the popping mechanism, we refer the reader to [5,6]. 

The goal of this paper is to discuss how to control the size of the
popped kernel of corn. In a recent paper by Hong and Both [7],
a method for controlling the size was developed based on an adiabatic
model of the popping mechanism. Because the expansion stops when the
vapor pressure becomes the same as the surrounding air pressure, all
that is needed to examine the change in size is the governing
equations immediately after the pericarp has broken. Therefore, the
problem of controlling the size reduces to that of interfacial
instability and pattern formation. In other words, we are simply
dealing with an interface advancing into a chamber. We will now
summarize the theory developed in [7].

The dynamic variable in the popcorn system will be the pressure $P$ inside 
the kernel at some time $t$. Because the popping mechanism is modeled as an 
adiabatic expansion, the pressure should obey the following law:
$$P(t)V(t)^\gamma = C_0 = {\mathrm {constant}},$$
where $V(t)$ is the volume of the kernel at $t$ and $\gamma$ is the usual 
ratio of the specific heats at a constant pressure and a constant volume, $\gamma = 
C_p / C_v$.   
Let $P_Y$ be the yield pressure at which the adiabatic expansion begins, and 
$V_0$ be the initial unpopped volume of the kernel. For water vapor, the 
value of $\gamma$ is known to be $1.3$ [8]. Substituting these values 
into the adiabatic equation, we find the following expression:
$$P_Y V_0^\gamma = C_0 = {\mathrm {constant}}, \eqno (1)$$
where $C_0$ can be determined by the yield pressure and the initial volume.  
For popcorn, we would use a value of $\gamma = 1.3$, the accepted value for water 
vapor [8].  
In treating the adiabatic expansion as an interfacial instability, the next 
step is to perform a linear stability analysis to determine whether the 
interface is stable against infinitesimal perturbations. The stability 
analysis was carried out by Both and Hong, where they found the interface was 
indeed stable. For more details on the stability analysis, see [7]. 

Knowing that the interface is stable, we are now ready to determine the size 
of the popped kernel when its internal pressure has reached that of the 
surroundings, $P_0$. If the kernel stops expanding at a set time $t_f$, then 
the internal pressure of the kernel, $P(t_f) = P_0$. From this pressure, we 
can use the adiabatic equation to determine the final volume of the popped 
kernel at time $t_f$,
$$ V(t_f) = \frac{C_0}{P_0}^{1/\gamma}. \eqno(2)$$
Recall from Eq.(1) that we have the constant $C_0$ in terms of the yield 
pressure $P_Y$ and the initial volume $V_0$. Substituting the value of 
$C_0$, we get a useful expression for the final volume,
$$V(t_f) = V_0\left(\frac{P_Y}{P_0}\right)^\frac{1}{\gamma}. \eqno (3)$$
Eq.(3) allows us to control final size of the popped kernel simply by 
modifying the pressure surrounding the kernel, $P_0$. From this equation we 
can estimate the yield pressure, $P_Y$ as well. We will use an assumption
that kernels of popcorn, both popped and unpopped, are approximately 
spherical. Now in reality, both the popped and unpopped kernel are far from 
spherical. However, this spherical assumption is reasonable since we are only 
interested in creating a simple theory to predict the ability to control 
the size, as opposed to a complex theory giving the exact change in size for 
an odd shaped kernel. For a typical kernel of popcorn, popped under normal 
atmospheric conditions, we estimate that the radius of the kernel 
increases by a factor of $4$ or more. Thus, the increase in the volume of 
a popped kernel is about a factor of $60$, or $V_f/V_0 \approx 60$. Now we 
can use Eq.(1) and Eq.(2) to get the following approximation for the ratio 
of the yield pressure to the pressure of the surroundings:
$$\frac{P_Y}{P_0} = \left(\frac{V_f}{V_0}\right)^\gamma \approx
60^{1.3} \approx 200. \eqno(4)$$
This allows us to estimate the yield pressure as
$$P_Y \approx 200 P_0,$$
which is about 200 times greater than atmospheric pressure. As noted in 
[7], the yield strength of the pericarp at its rupture point is about 
the same or greater than polyethylene (LDPE) at room temperature [9]. 
Now using, the equations above, we want to derive an expression allowing the 
size of the popped kernel to be controlled by adjusting a simple parameter of 
the system. We first define the variable $\Gamma$ to be the ratio of the 
final volume $V_f$ to the initial volume $V_0$, 
$$\Gamma(P_0) = \frac{V(t_f)}{V_0},$$
which we will call the volume expansion rate. Using Eq.(2) and Eq.(4), we 
get the following expression for $\Gamma $:
$$\Gamma(P_0) =
\left(\frac{1}{V_0}\right)\left(\frac{C_0}{P_0}\right)^{\frac{1}{\gamma}}.
\eqno (5)$$
From Eq.(5), it is obvious that $P_0$, the surrounding pressure, is
the parameter that will control the increase in size between a
kernel's initial and final states. According to this simple
thermodynamic theory, the lower the pressure that surrounds the
kernel, the larger the increase in volume of the kernel when it
pops. Hong and Both [7] took the adiabatic analysis even
further and predicted how much the volume of a popped kernel would 
change with a fixed deviation from atmospheric pressure. In this
chapter, we are more interested in testing the qualitative prediction,
that one can increase the size of a piece of popcorn by lowering the
surrounding pressure, $P_0$. Therefore, we will not discuss the
details of the quantitative analysis, but instead refer the reader to
[7]. Now we present the experiment, the results, and the
comparison with the theory.

\section{The Popcorn Experiment}
There are many ways that one can choose to pop popcorn. The
conventional way is to pop it is on the stove in a pan with some
oil. Other popular methods of popping include using the microwave oven
and the air popper. In choosing our method of popping to test the
theory, it was very important to think about the moisture content of
the corn and how the heat is distributed to the kernels. According to
the literature, moisture content and the distribution of heat among
the kernels play a very important role in popping
[3,4,6].  Popcorn pops best when its moisture content
is somewhere between 11\% and 14\% percent. If the kernels are dry, the
internal pressure does not get high enough to break through the
pericarp. For kernels with too much moisture, the pericarp becomes
soft and cracks prematurely, before the starch has reached its
jelly-like state. Therefore it is best to pick a popping method that
works best for a wide range of moisture

\begin{figure}
\begin{center}
\epsfig{file=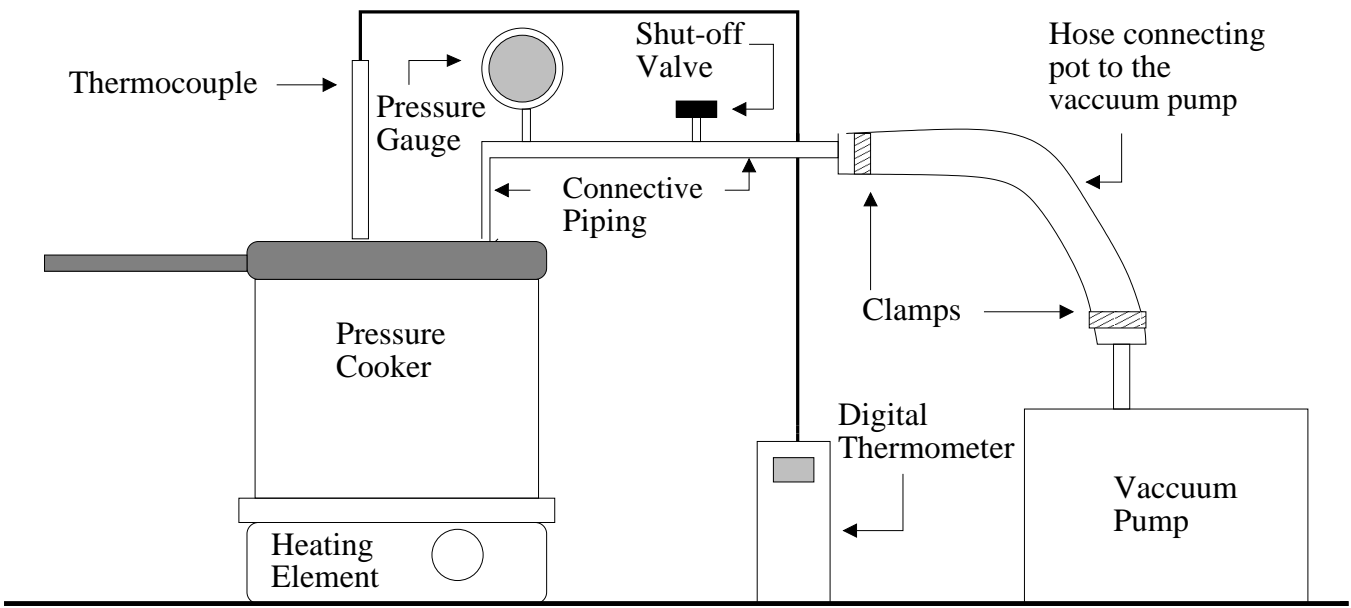,height=8cm,angle=90,clip=}
\end{center}
\caption[A schematic of the apparatus used for popping kernels in a pot.]{A 
schematic of the apparatus used to pop the kernels in a pot. The apparatus
includes a heating element to heat the pot and a vacuum pump to lower
the pressure.}
\end{figure}

\noindent levels. The same is true for
heating. If the kernel is not heated evenly, the results will be a
burnt unpopped kernel, or a semi-popped kernel that is burnt on one
side. It is also important to heat the kernels rapidly, so as to
``shock'' the kernels into an adiabatic expansion, not giving them time
to adjust or equilibrate to the temperature of the surroundings. If
the kernels are cooked gradually, starting at a low heat, the pericarp
will once again soften and crack, allowing the moisture to escape out
slowly rather than popping out rapidly.  Since the moisture content of
most brands of popcorn is not readily available, we arbitrarily chose
to use Jolly-Time popcorn for all of our experiments. We chose the
following two methods of popping to test the predictions of the
theory: popping with oil in a pan and dry popping in a mesh bottomed pot.   
In the future, designing an apparatus that uses microwaves to heat the 
kernels would be another useful way of testing this theory.  Not only is the 
microwave oven the best way to evenly heat your kernels, it is the most 
efficient and successful method used by the industry to produce popcorn.  Direct 
comparisons could then be made between the experimental results and those 
reported by popcorn manufaturers.  However, in this particular paper, we were only
interested in determining whether this thermodynamic model is qualitatively correct in 
predicting experimental results.  For the purposes of this paper, the two chosen 
methods of heating the kernels were sufficient.  

Our experimental method of popping the popcorn in a pot is one that is familiar to 
industry as well as people in the common household. Many people who make popcorn at 
home pop it in a pan with some oil on the bottom. We designed an apparatus, 
shown in Fig.(1), to see if lowering the surrounding pressure in the pot would yield
popped kernels with an increased volume. We started with a standard
pressure cooker and altered it to include a vacuum pump used to remove
the air and hence, lower the pressure in the pot. A thermocouple
connected to a thermometer and a pressure gauge were included to allow
measurements of the temperature and pressure inside the pot. The
temperature of the pot was controlled
with a heating pad attached to the bottom of the pot. 
A shut-off valve was built into the apparatus, allowing
us to break the vacuum seal once the experiment was completed.  The
procedure for conducting the experiment was fairly simple and very
systematic. The first set of data was gathered for popcorn popped in
the bottom of the pot with oil. Since most of the experiments
conducted in the literature [3,4] worked with $20$ g
samples of unpopped kernels, we decided to use the same sample size
for our experiments. This sample size resulted in about 135--155 kernels,
depending on their size and shape. All samples were kept in sealed
bags before use, to prevent them from drying out. The first step was
to heat the empty pot to a temperature of $150^\circ$ Celsius.
Meanwhile, the kernels were mixed with the oil while the pot heated
up. When the proper temperature was reached, the oil soaked kernels
were put inside the pot, which was then sealed shut. Two runs were
done for each set of data, one at atmospheric pressure and one at a pressure 
reduced to $P \approx 1/30$ atm. For the runs conducted at atmospheric pressure, 
the kernels were left in the 
sealed pot until the popping was completed. To conduct a run at low pressure, 
the vacuum pump had to be turned on after the pot was sealed shut with the 
kernels inside. When the kernels popped, water vapor is released into the pot, 
causing a slight increase in pressure. Therefore, it was necessary to keep 
the vacuum pump running throughout the completion of the popping in order to 
remove this excess water vapor and maintain a low pressure. Once two runs were 
completed, a comparison could be made between the two samples to see if the volume 
of the popped kernels had increased under low pressure. 

The second method of popping we chose was dry popping the kernels on a 
steel mesh placed in the bottom of the pot. Oil is used 
for making popcorn because it coats the kernels and allows for an even 
distribution of heat over the entire pericarp, or shell. However, we were 
curious to see if the coating of oil that surrounded each kernel was limiting 
the size increase of the popcorn as it expanded. Therefore, we chose to 
eliminate the oil and pop it dry. The mesh was needed to ensure that 
the heat was more evenly distributed and to keep the kernels away from the 
bottom of the pan, where they would burn. The procedure was conducted the 
same as the oil popping method, the only difference being the placing of dry 
kernels on the mesh, as opposed to the oil soaked mixture used previously.    

\section{Experimental Results}
Numerous runs were conducted for both methods of popping. A sample of some results 
and comparisons for both methods of popping, are shown in Figs.(2--4) as well as 
Tables 1 and 2. These tables are a small sample of
results gathered from individual trials of each experiment. The
results in Figs.(2--4) are the average taken over the numerous set
of trials we conducted. We made comparisons for the three parameters
$\sigma$, $\pi$, and $\omega$ to see if they were altered by popping
at a lower pressure. The first graph, Fig.(2), shows that lowering
the pressure more than doubles the value of $\sigma$, the expansion volume. 
This was the case, regardless of whether 
we used oil or not.  In Fig.(3), one can see that lowering the
pressure increased the flake size, $\pi$, by approximately a factor of
two as well. At first glance, these results can be deceiving.  One
might be led to believe that 

\begin{table}
\begin{center}
\begin{tabular}{|c|c|c|c|} 
\hline
\multicolumn{4}{|c|}{Popped in Oil under Atmospheric Pressure}\\ \hline
Trial& Flake Size & Expansion Volume & Unpopped Kernels \\
&  $\pi$ (cm$^3$) & $\sigma$ (cm$^3$/g) & $\omega$ (\%)\\ \hline
1 & 2.76 & 17.5 & 15.3 \\
2 & 2.41 & 10.0 & 40.7 \\
3 & 1.99 & 11.3 & 23.1 \\
4 & 2.35 & 10.0 & 41.8 \\ 
5 & 2.48 & 12.5 & 29.9 \\ \hline
\end{tabular}
\vskip .5 true cm
\begin{tabular}{|c|c|c|c|} 
\hline
\multicolumn{4}{|c|}{Popped in Oil Under Reduced Pressure}\\ \hline
Trial& Flake Size & Expansion Volume & Unpopped Kernels \\
&  $\pi$ (cm$^3$) & $\sigma$ (cm$^3$/g) & $\omega$ (\%)\\ \hline
1 & 3.75 & 27.5 & 3.4 \\
2 & 3.85 & 27.5 & 4.0 \\
3 & 3.62 & 25.0 & 5.5 \\
4 & 3.79 & 25.0 & 8.3 \\ 
5 & 4.14 & 27.5 & 3.8 \\ \hline
\end{tabular}
\caption [A sample of average values per trial of $\pi$, $\sigma$ and $\omega$ 
for corn popped in oil.]{A sample of average values per trial of $\pi$, 
$\sigma$ and $\omega$ for popcorn popped in oil. The trials on top were 
conducted at regular pressure, while those on the bottom were done at a 
reduced pressure of $P \approx 1/30$ atm.}
\end{center}
\vskip 1.2 true cm
\begin{center}
\begin{tabular}{|c|c|c|c|} 
\hline
\multicolumn{4}{|c|}{Popped on Mesh under Atmospheric Pressure}\\ \hline
Trial& Flake Size & Expansion Volume & Unpopped Kernels \\
&  $\pi$ (cm$^3$) & $\sigma$ (cm$^3$/g) & $\omega$ (\%)\\ \hline
1 & 2.34 & 12.5 & 25.7 \\
2 & 2.86 & 15.0 & 27.1 \\
3 & 2.15 & 10.0 & 38.2 \\
4 & 2.17 & 12.5 & 24.8 \\ 
5 & 2.35 & 12.5 & 27.9 \\ \hline
\end{tabular}
\vskip .5 true cm
\begin{tabular}{|c|c|c|c|} 
\hline
\multicolumn{4}{|c|}{Popped on Mesh under Reduced Pressure}\\ \hline
Trial& Flake Size & Expansion Volume & Unpopped Kernels \\
&  $\pi$ (cm$^3$) & $\sigma$ (cm$^3$/g) & $\omega$ (\%)\\ \hline
1 & 3.96 & 27.5 & 3.5 \\
2 & 4.10 & 27.5 & 6.3 \\
3 & 3.90 & 27.5 & 2.1 \\
4 & 4.01 & 27.5 & 5.8 \\ 
5 & 3.23 & 22.5 & 8.5 \\ \hline
\end{tabular}
\caption [A sample of average values per trial of $\pi$, $\sigma$ and $\omega$ 
for popcorn dry popped on a mesh.]{A sample of average values per trial of $\pi$, 
$\sigma$ and $\omega$ for corn dry popped on a mesh. The trials on top were 
conducted at regular pressure, while those on the bottom were done at a 
reduced pressure of $P \approx 1/30$ atm.}
\end{center}
\end{table}

\begin{figure}
\begin{center}
\vskip .5 true cm
\scalebox{.5}{
\includegraphics[100,0][700,700]{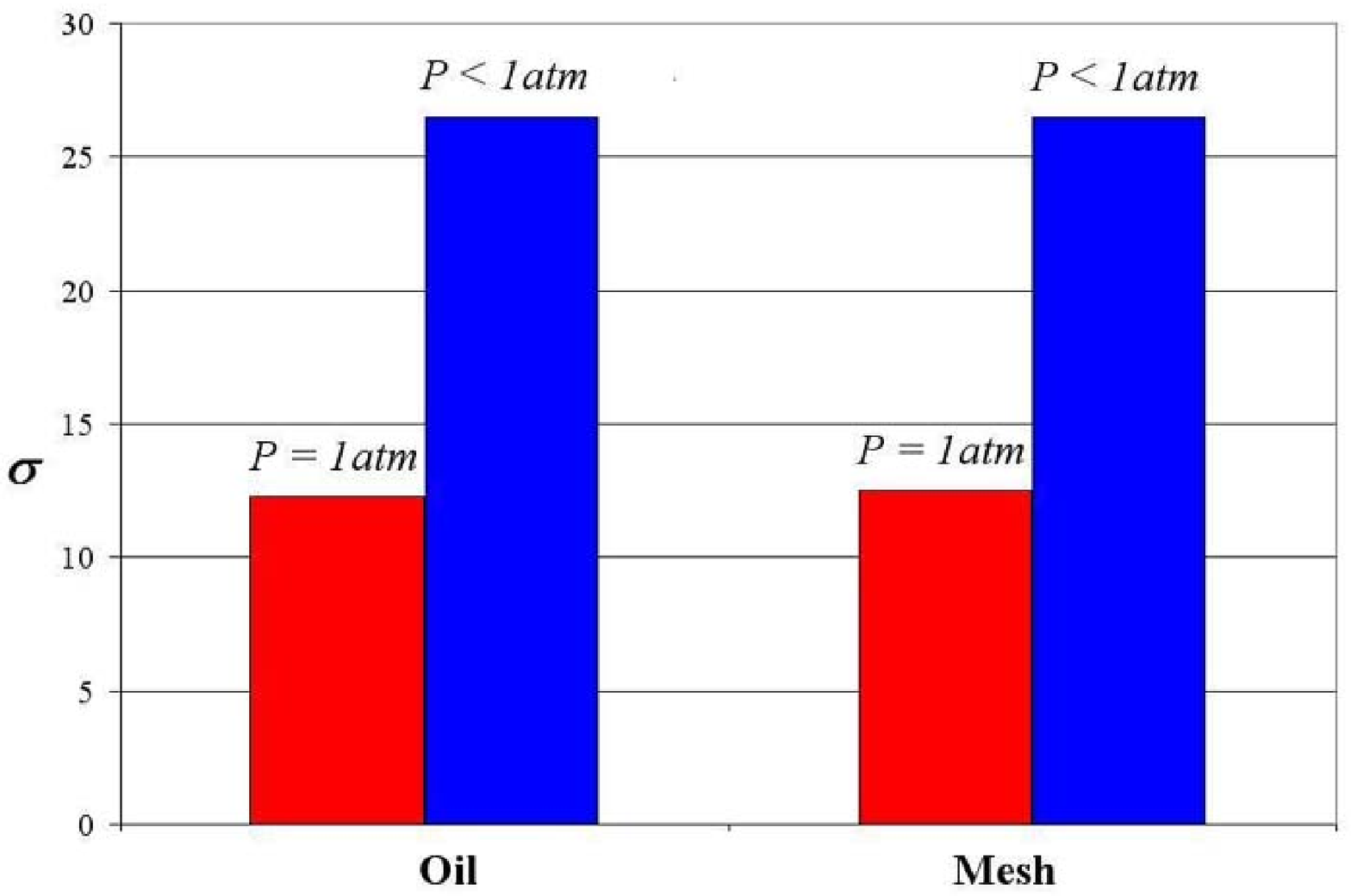}}

\end{center}
\caption[A comparison of $\sigma$ between kernels popped at $P 
\approx 1$ atm and $P \approx 1/30$ atm in a pot.]{A comparison of $\sigma$ 
between kernels popped at regular atmospheric pressure and those
popped at a reduced pressure of $P \approx 1/30$ atm. This graph shows
the comparison of $\sigma$ for both the oil and mesh popping.}
\end{figure}

\begin{figure}
\begin{center}
\vskip .5 true cm
\scalebox{.5}{
\includegraphics[100,0][700,700]{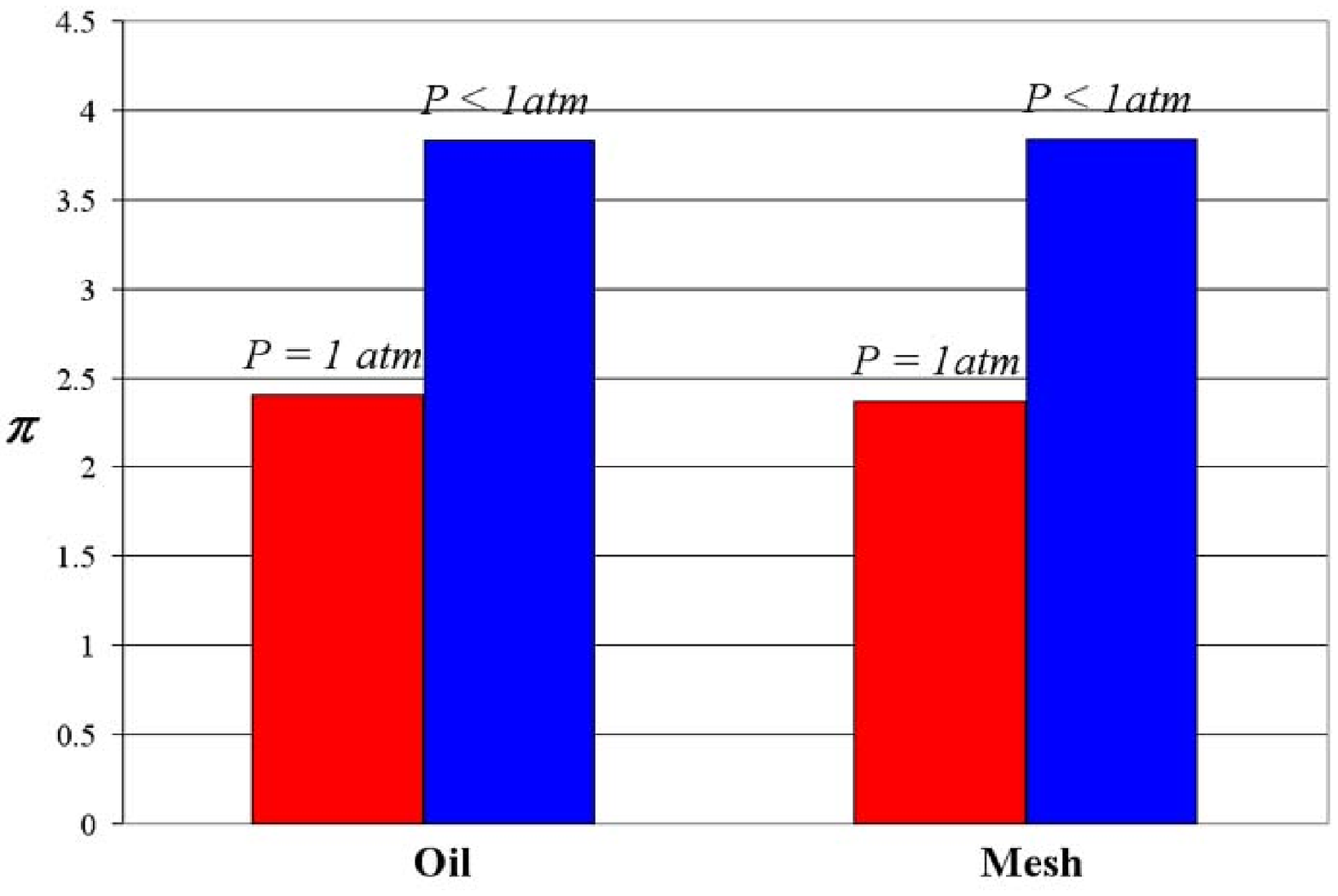}}
\end{center}
\caption[A comparison of $\pi$ between kernels popped at $P 
\approx 1$ atm and $P \approx 1/30$ atm in a pot.]{A comparison of $\pi$ 
between kernels popped at regular atmospheric pressure and those
popped at a reduced pressure of $P \approx 1/30$ atm. This graph shows
the comparison of $\pi$ for both the oil and mesh popping.}
\end{figure}

\begin{figure}
\begin{center}
\vskip .5 true cm
\scalebox{.4}{
\includegraphics[250,0][700,700]{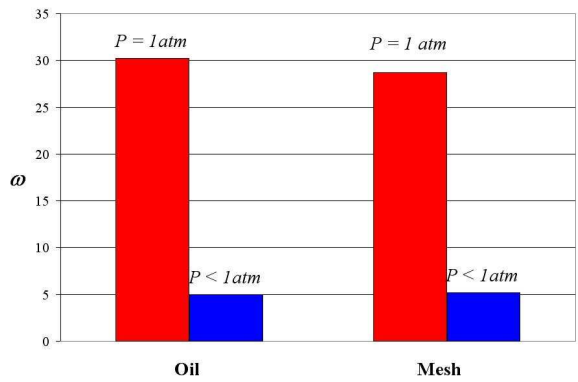}}
\end{center}
\caption[A comparison of unpopped kernels, $\omega$, between kernels 
popped at $P \approx 1$ atm and $P \approx 1/30$ atm in a pot]{A
comparison of unpopped kernels, $\omega$, between kernels popped at
regular atmospheric pressure and those popped at a reduced pressure of
$P \approx 1/30$ atm.  The results are shown for both the runs using
oil and the mesh.}
\end{figure}

\noindent since the volume of the popcorn was
approximately doubled, we have produced a popped kernel that looks
twice as big. Recall, though, that our calculations were done using a
spherical approximation for the volume of a 
kernel of popcorn. Using the definition for the volume of a 
sphere, $V = \frac{4}{3} \pi r^3$, one can see that if the volume is
doubled, the radius is only increased by a factor of $2^{\frac{1}{3}}
\approx 1.2$. This is an increase in the radius, but not one that can
be observed with the naked eye, specifically, in a popcorn sample
containing a variety of shapes and sizes. However, the volume has
significantly increased, which can be seen clearly from our results. Under
regular popping, 20 g of unpopped kernels yields {\it one} 8 oz cup of
popcorn, while 20 g popped in a low pressure environment yields {\it
two} 8 oz cups. The predicted increase in size by the theory was
qualitatively confirmed by our experiments.
Another interesting result that we observed was an extreme decrease in the 
value of $\omega$, the percentage of unpopped kernels in a sample. 
As shown in Fig.(4), the amount of wasted kernels decreases by about a 
factor of 5. This means that popping under a reduced pressure is, in fact, a 
more efficient method of producing popcorn. This is a result that could be 
very useful for the popcorn producing industry. Means of production could be 
made more efficient, reducing the cost acquired by the producers and hence, 
increasing overall profits.

\section{Conclusion}

It is clear from the presented results that the qualitative predictions of 
[7] are correct. The popping process in a pot, whether using oil or dry 
popping with the mesh on the bottom, displayed an increase in the size
of the kernels due to the reduced pressure of the
surroundings. Popping under reduced pressure also seemed to be a more
efficient method of making popcorn, shown by the decrease in unpopped
kernels that were observed. The results seem to indicate that
regardless of the popping method, it may be possible, on average, to
double the volume of popped kernels. Recall that the maximum reported
values for the volume expansion, $\sigma$ and the flake size, $\pi$,
are reported to be approximately $36-40$ cm$^3$/g and $5-7$
cm$^3$/kernel, respectively [1,2]. It is important to note
that even though we have demonstrated an increase in $\sigma$ and
$\pi$ that occurs when the pressure is reduce, our maximum values of
$27.5$ cm$^3$/g and $4.01$ cm$^3$/kernel are still smaller then those
previously reported [1,2]. However, those previously
reported values were obtained without lowering the pressure, in an
air popper or a microwave. So if the pattern observed in our data was
to remain true, an air popper or a microwave oven placed in a reduced 
pressure environment, could possibly produce double the values of the 
maximums currently observed in the literature today. I would expect at 
least some increase in the size due to the lowering of the pressure.  It is 
important to keep in mind that the expansion of the starch in the popped kernel 
does have a physical limit.  Doubling the results obtained by the popcorn 
industry may be well beyond the physical size limitiations of the start.  However, 
it would be interesting to see if results that are better than those obtained by
industrial methods can be produced with a more efficient apparatus, such as the 
microwave oven.    

For this idea to catch the interest of popcorn producers, an attempt must be 
made to obtain results that outdo the maximum values reported by industry. This 
is a design problem that can
be worked out by an engineer and then used to further test this theory. We
have clearly shown that the size can be controlled by changing the
surrounding pressure. Now it is just a matter of perfecting the
mechanism that uses this premise. It is also important to note, that
reducing the pressure lowered the unpopped number of kernels by a
factor of 5. This should be of real interest to the popcorn
producers. By reducing the number of wasted kernels, manufacturers
will get more popcorn for the same cost of supplies. This could be a
very effective way to cut down on costs, and hence, make more
profit. More work needs to be done to see if the effects produced by
lowering the pressure, can be magnified even further.  However, we
have demonstrated beyond a doubt, that the surrounding pressure is a
very important parameter in the production of popcorn.

\vskip .8 true cm   

\Large 
\noindent \bf Acknowledgements
\normalsize
\normalfont

\vskip .4 true cm 

I would like to acknowledge Dr. Daniel C. Hong for all his work on this project. Without 
his ingenuity and creative thinking, this project would have never come to fruition. 
Unfortunately Dr. Hong passed away in 2002 and was unable to see this process to 
completion.  Dr. Hong was a great scientist, colleague, mentor, and friend.  The scientific
world has lost a wonderful scientist and a fantastic theoretical mind.  His presence in the 
physics community will surely be missed.  I would also like to thank Joe Zelinski at Lehigh 
University for his help in designing and constructing the apparatus used for this 
experiment.  I would also like to acknowledge the physics department at Lehigh University 
for providing the funds necessary to construct the apparatus.

\vskip .8 true cm 

\noindent [1] A. A. Mohamed, R. B. Ashman, and A. W. Kirleis, \it{Journal of Food Science},
\normalfont \bf{58}, \normalfont No.2, 342(1993).

\vskip .2 true cm

\noindent [2] R. C. Hoseny, K. Zeleznak, and A. Abdelrahman, \it{Journal of Cereal Science},
\normalfont \bf{1}, \normalfont 43(1983).

\vskip .2 true cm

\noindent [3] T. H. Roshdy, K. Harakawa, and H. Daun, \it{Journal of Food Science},
\normalfont \bf{49}, \normalfont 1412(1984).

\vskip .2 true cm

\noindent [4] C. G. Haugh, R. M. Lien, R. E. Hanes, and R. B. Ashman, 
\it{Trans. Am. Soc. Agric. Eng.}, \normalfont \bf{19}, \normalfont 
168(1976).

\vskip .2 true cm

\noindent [5] Robert G. Hunt, \it{The Physics Teacher}, \normalfont April, 
230(1991).

\vskip .2 true cm

\noindent [6] W. J. Silva et al, \it{Nature}, \normalfont April, 
No. 6419,362(1993).

\vskip .2 true cm

\noindent [7] D. C. Hong and Joseph A. Both, \it{Physica A}, \normalfont
\bf{289}, \normalfont 557(2001).

\vskip .2 true cm

\noindent [8] \it {CRC Handbook of Physics and Chemistry}, \normalfont
CRC Press,Boca Raton, Florida,D-172,(1985).

\vskip .2 true cm

\noindent [9] N. E. Dowling, \it{Mechanical Behavior of Materials},
\normalfont Prentice Hall, Englewood Cliffs, New Jersey, 158(1993).

\end{document}